\documentclass[aps,prl,twocolumn,amsmath,letterpaper]{revtex4}

\usepackage{graphicx}           
\usepackage{hyperref}
\usepackage{color}
\newcommand{\iea}[0]{{\it et al.}}

\newcommand{\ket}[1]{\left| #1\right\rangle}

\newcommand{\eeqref}[1]{Eq.~(\ref{#1})}

\begin{document}
\title{Photons as quasi-charged particles }

\author{K.-P. Marzlin, J\"urgen Appel, A. I. Lvovsky}

\affiliation{Institute for Quantum Information Science, University of
  Calgary, Calgary, Alberta T2N 1N4,
  Canada}

\date{\today}

\begin{abstract}
{
The Schr\"odinger motion of a charged quantum particle in an electromagnetic
potential can be simulated by the paraxial dynamics of photons 
propagating through a spatially inhomogeneous medium. 
The inhomogeneity induces geometric effects that generate an artificial
vector potential to which signal photons are coupled. 
This phenomenon can be implemented with slow light propagating through
an a gas of double-$\Lambda$ atoms in an electromagnetically-induced 
transparency setting with spatially varied control fields.
It can lead to a reduced dispersion of signal photons and 
a topological phase shift of Aharonov-Bohm type.
}
\end{abstract}

\maketitle

{\em Introduction.---} \label{sec:Introduction}
It is known since the ground-breaking work of Berry on geometric phases
\cite{berry:1984} that artificial gauge potentials can be induced if the
spatial dynamics of a system that obeys a wave equation is confined in
a certain way. For instance, if the internal Hamiltonian of neutral atoms
contains an energy barrier but the spin eigenstates are spatially varying,
gauge field dynamics can be induced \cite{STIRAP}.
In the limit of ray optics, moving atomic ensembles could simulate the propagation
of light around a black hole or generate topological
phase factors of the Aharonov-Bohm type \cite{PRA60:4301},
and inhomogeneous dielectric media could generally
exhibit geometric effects such as an optical spin-Hall effect
and the optical Magnus force \cite{Magnus}.

In this paper, we propose to use electromagnetically induced transparency (EIT)
to generate an artificial vector potential
for the paraxial dynamics of signal photons that simulates \emph{quantum dynamics} of 
charged particles in a static electromagnetic field. Not only the ray of light but
also its mode structure is affected, resulting in a paraxial wave equation
that is equivalent to the Schr\"odinger equation for charged particles.
Furthermore, the form of the artificial vector potential can be easily
controlled through spatial variations in the control fields. We suggest
configurations that generate homogeneous quasi-electric and magnetic
fields as well as a vector potential of Aharonov-Bohm type.

Although the treatment in this paper is based on EIT, the effect presented here is more general: 
it will occur in any medium that supports a set of discrete eigenmodes for a propagating signal 
fields with different indices of refraction. If the parameters governing these eigenmodes vary 
in space, the signal modes will adiabatically follow, acquiring geometric phases that affect their 
paraxial dynamics.

\begin{figure}[tbp]
  \centering
  \includegraphics[width=0.5\columnwidth,keepaspectratio]{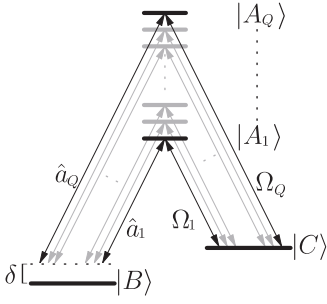}
  \caption{Multi $\Lambda$-system: $Q$ excited states
    $\ket{A_q}$ are each coupled
    by a classical control field $\Omega_q$ to
    the ground state $\ket{C}$ and by a quantized field $\hat a_q$
    with detuning $\delta$ to state $\ket{B}$.}
  \label{fig:multilambda1}
\end{figure}

{\em Review of EIT with multi-$\Lambda$ atoms.---}
\label{sec:Discrete_Mode_Field}
The effect takes place in an atomic
multi-$\Lambda$ system, in which two ground states are coupled to $Q$
excited states by $Q$ pairs of control ($\Omega_q$) and signal ($\hat a_q$) fields
(Fig.~\ref{fig:multilambda1}).
An experimentally relevant example of such system is the fundamental D1
transition in atomic rubidium, where both the ground and excited levels
are split into two hyperfine sublevels \cite{vewinger}.
We assume that the detunings are small so each
signal field $\hat a_q$ interacts only with the respective
transition $ \ket{B} \leftrightarrow \ket{A_q}$ with the associated atomic
operator $\hat{\sigma}_{B,A_q}$ and vacuum Rabi frequency $g_q$. In this case, the
paraxial wave equation for each signal mode can be cast
into the form
\begin{equation} \label{parax}
  \Bigl( \frac{\partial}{\partial t} +
  c \frac{\partial}{\partial z}
  - \frac{i c}{2k} \Delta_\perp \Bigr)
  \hat{a}_q = i N g_q \hat{\sigma}_{B,A_q}\; ,
\end{equation}
where the wave propagates along the $z$ axis, $\Delta_\perp =
\partial^2_x+\partial^2_y$, $N$ is the number of atoms and $k$ is the
wavevector which we assume approximately independent of $q$. In Ref.~\cite{appel:013804} we
have constructed a unitary transformation
\begin{equation} \label{W}\hat a_q=\sum_{s=1}^Q W_{qs} \, \hat b_{s}
\end{equation}
that maps the original field modes $a_q$ to a new set of modes $\hat b_q$,
such that one and
only one of the new modes,
\begin{equation} \label{bQ}
  \hat b_Q = \sum_{q=1}^Q R_q^*  \hat a_q,
\end{equation}
(where $R_q \equiv \Omega_q/(g_q \Omega_\perp)$ and
$ \Omega_\bot \equiv \sqrt{\sum_{q=1}^{Q} |\Omega_q/g_q|^2}$ depend on the control fields)
couples only to an atomic dark state and
experiences EIT \cite{appel:013804,Liu06,Moiseev}. All other superpositions of field modes are absorbed.
This transformation is given explicitly by $ W_{qq'} = \gamma w_q w_{q'}^* -\delta_{qq'}$,
with $\gamma= R_Q +1$ and $w_q= \gamma^{-1} (\delta_{Qq} +   R_q )$.

The EIT mode $\hat b_Q$ interacts with the multi-$\Lambda$ atoms in the same
fashion as does the signal field in a regular 3-level system. While
propagating through the EIT medium, it gives rise to a dark-state polariton
associated with zero interaction energy \cite{fleischauer}. All other modes
couple to atomic states whose
energy levels are Stark-shifted by the interaction with either the
pump field or the other signal modes $\hat b_q$ ($q\neq Q$). The resulting energy
gap guarantees that, if the amplitudes and phases of the control fields are
slowly changed, the composition of the dark-state
polariton, and hence the EIT mode $\hat b_Q$, will adiabatically follow.
It has been proposed \cite{appel:013804} and experimentally demonstrated
\cite{vewinger} that a variation in time of the control fields
can therefore be used to adiabatically transfer optical states between
signal modes. In this paper, we
focus on \emph{spatial} propagation of the EIT mode under control fields that
are constant in time, but varied in space.

{\em Derivation of the gauge potential.---}
We proceed by expressing Eq.~(\ref{parax}) in terms of the new signal modes $\hat{b}_q$.
Employing the vector notation
$\vec{a} = \{ \hat{a}_1 , \cdots , \hat{a}_Q \}$
and $\vec{\sigma}_{B,A} = \{ g_1 \hat{\sigma}_{B,A_1}, \cdots ,
g_Q \hat{\sigma}_{B,A_Q} \}$
we get
\begin{equation}
  \label{eq:33a}
  \Bigl( \frac{\partial}{\partial t} +
  c \frac{\partial}{\partial z}
  - \frac{i c}{2k} \Delta_\perp \Bigr)
  \tensor{W} \vec{\hat b} = i N \vec{\sigma}_{B,A}.
\end{equation}
Throughout the paper, the double arrow denotes a $Q\times Q$ matrix. 
Because $\tensor W$ depends on space and time, the differential operators
have to be applied to both $\tensor {W}$ and $\vec{\hat b}$.
As a result, transformation (\ref{W}) brings about additional terms into the
equation of motion, that can be written in form of a minimal coupling scheme
by introducing the Hermitian gauge field
\begin{equation} \label{AW}
  \tensor {A}_i \equiv i \tensor {W}^\dagger \partial_i \tensor {W} ,
\end{equation}
where $i=t,x,y,z$.
We multiply both sides of \eeqref{eq:33a} by $\tensor {W}^{\dagger}$ and exploit the unitarity of $\tensor {W}$
to show that $\partial_i \tensor {W}^\dagger = -  \tensor {W}^\dagger
 (\partial_i \tensor {W}) \tensor {W}^\dagger$ from which it follows that
$- \tensor {W}^\dagger  \partial_i^2 \tensor {W} = \tensor {A}_i^2 + i \partial_i \tensor {A}_i$.
The dynamic equation for the $\hat b$ modes can then be written as
\begin{eqnarray}
  \left ( i \partial_t + \tensor A_t \right ) \vec{\hat b} &=&
 -\Big ( i c \partial_z + c \tensor A_z \Big ) \vec{\hat b}
\\ & &
  + \frac{c}{2k} ( -i \nabla_\perp - \tensor A_\perp )^2  \vec{\hat b}
  -\tensor {W}^\dagger N \vec{\sigma}_{BA}
\nonumber\end{eqnarray}
with $\nabla_\perp = (\partial_x, \partial_y)$. This equation
has the structure of a 2+2 dimensional field theory with minimal coupling.

Under the assumption that the
control fields do not depend on $t$ and $z$ we can make a temporal
Fourier transformation of the slowly varying amplitudes, which results
in the paraxial wave equation
\begin{eqnarray} \label{parbq}
  i \partial_z  \vec{\hat b}(\delta) &=& \left ( -\frac{\delta }{c}
  + \frac{1}{2k} ( -i \nabla_\perp - \tensor A_\perp )^2 \right )  \vec{\hat b}(\delta)
\nonumber \\ & &
  -\tensor {W}^\dagger \frac{ N }{c} \vec{\sigma}_{BA}(\delta).
\end{eqnarray}

The gauge potential is given explicitly by
\begin{equation}
  \tensor A_\perp = i \sum_{q=1}^Q R_q^* (\nabla_\perp R_q ) \vec{w} \vec{w}^\dagger
    -i\gamma  (\nabla_\perp \vec{w})  \vec{w}^\dagger
    + i \gamma^* \vec{w} \nabla_\perp \vec{w}^\dagger \; .
\end{equation}
The full matrix $\tensor A_\perp$ is a pure gauge: it has emerged solely as a
consequence of the unitary transformation (\ref{W}), which reflects our choice
to describe the system in terms of the new modes $\hat b_q$ rather than the
original modes $\hat a_q$. However, this choice is motivated by the fact that
the EIT mode $\hat b_Q$ is the only mode that is not absorbed. Absorption of other 
modes $\hat b_q$ (with $q\ne Q$) means that the index of refraction for these 
modes has a significant imaginary part. 
This separates the EIT mode $\hat b_Q$ from other $b$-modes and ensures that it will 
adiabatically follow variations of the control fields. Therefore,
when analyzing the evolution of $\hat b_Q$, we can neglect the off-diagonal
terms in the matrix $( -i \nabla_\perp - \tensor A_\perp )^2$ in \eeqref{parbq} and
write
\begin{eqnarray}
  i \partial_z  \hat b_Q(\delta) &=&
  -(\tensor {W}^\dagger \frac{ N}{c} \vec{\sigma}_{BA})_Q (\delta)
  -  \frac{\delta }{c} \hat b_Q(\delta)
 \label{gaugeParax}\\ & &
  + \frac{1}{2k} ( -i \nabla_\perp - A_\perp )_{Qq}
  ( -i \nabla_\perp - A_\perp )_{qQ}
  \hat b_Q(\delta).
\nonumber\end{eqnarray}
This equation does not include the whole matrix $\tensor A_\perp$. Consequently,
this potential no longer acts like a pure gauge but attains physical significance
in determining the spatial dynamics of the EIT mode.

The first term
on the right-hand side of Eq.~(\ref{gaugeParax}), responsible for the interaction of the light field with the EIT medium, takes the same form
as the susceptibility of EIT in a single $\Lambda$-system.
Neglecting decoherence, we can write it as \cite{appel:013804}
$  (\tensor {W}^\dagger \frac{ N}{c} \vec{\sigma}_{BA})_Q (\delta)
  =  \frac{ \delta}{v_\text{EIT}} \hat b_Q$,
with the EIT group velocity 
$v_\text{EIT} = c \Omega^2 /(N g^2)$.
Note that $v_\text{EIT}$ depends on the spatial position because $\Omega$ does.
This transforms Eq.~(\ref{gaugeParax}) to
\begin{equation} 
  i \partial_z  \hat b_Q = \left (
 \frac{1}{2k} ( -i \nabla_\perp - A_{QQ})^2 - \frac{ \delta}{v_\text{EIT}}
 + \frac{\Phi}{2k} \right) \hat b_Q
\label{gaugeParax2}\end{equation} 
with \begin{eqnarray}
  A_{QQ} &=& i \sum_{q=1}^Q R_q^* \nabla_\perp R_q
    = -  \sum_{q=1}^Q |R_q|^2 \nabla_\perp \text{Arg}(R_q), \nonumber
\\
  \Phi &\equiv& \sum_{q=1}^{Q-1} |(A_\perp )_{Qq}|^2=
  -A_{QQ}^2  + \sum_{q=1}^{Q} |\nabla_\perp R_q|^2
  \label{AqQsum}
\end{eqnarray}
being, respectively, the ``quasi-vector'' and ``quasi-scalar'' potentials.

We see that the paraxial spatial evolution of the EIT signal mode is governed
by the equation that is identical (up to coefficients) to the Schr\"odinger
equation of a charged particle in an electromagnetic field. This is the main
result of this work. By arranging the control field in a certain
configuration, one can control the spatial propagation of the signal mode
through the EIT medium.

Some steering of the EIT mode is possible even in a single-$\Lambda$ system by
affecting the term ${ \delta}/v_\text{EIT}$ in \eeqref{gaugeParax2}, which
results in nonuniform refraction for this mode
\cite{waveguiding,SternGerlachEIT}. The action of quasi-gauge fields
(\ref{AqQsum}) is fundamentally different: deflection of the signal field
occurs not due to refraction (the refraction index on resonance is 1), but due
to adiabatic following.

{\em The case of two control fields: homogeneous electric and magnetic
quasi-fields.---}
Of particular practical importance is the simplest non-trivial case
with $Q=2$. We parametrize the control fields by writing
$R_{1,2} = \sqrt{1/2 \pm R}\, e^{i (\phi \pm\theta)}$. The corresponding Rabi
frequencies are then $\Omega_i = h(x,y) \, g_iR_i$, with $h(x,y)$ being an
arbitrary common prefactor.
This parametrization yields the gauge potentials
\begin{eqnarray}
  A_{QQ} &=& - \nabla_\perp \phi - 2 R \nabla_\perp \theta;
\label{AQQ2} \\
  \Phi &=&
  \frac{ (\nabla_\perp R)^2}{1-4 R^2} + (\nabla_\perp \theta)^2 (1-4R^2).\nonumber
\end{eqnarray}
Similarly to usual electrodynamics we can use a gauge transformation \cite{GaugeFootnote},
$A_{QQ}' = A_{QQ} + \nabla_\perp f$, to eliminate the term
$\nabla_\perp \phi$ from Eq.~(\ref{AQQ2}). The common phase $\phi$
of the control fields therefore does not contribute and can be set to zero.

A simple way to generate a term that corresponds to a one-dimensional
scalar potential $V(x)$ for a Schr\"odinger particle is to choose $R=0$ and
$\theta = \int_0^x \text{d}x'\; \sqrt{2 k V(x')}$
This choice of control
fields leads to $A_{QQ}=0$ and $\Phi = 2 k V(x)$.

\begin{figure}[htbp]
  \centering
  \includegraphics[width=0.95\columnwidth,keepaspectratio]{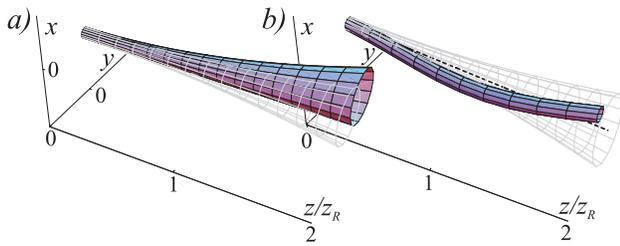}
  \caption{\label{EBfieldFig}Paraxial propagation of a signal beam over twice the Rayleigh length
  in the presence (solid) and absence (grey) of a
  constant (a) electric field along the $x$ axis and (b) magnetic field along $z$. The dashed
  line represents the center of the grey beam. The effect of the fields is somewhat exaggerated.}
\end{figure}

For the special case
of a constant electric quasi-field along the $x$ axis, $V(x) =  - F x$ and subsequently
\begin{equation} \label{thetaEF} \theta = - \sqrt{4 k F |x|^3/3},\end{equation}
where $x<0$ is assumed for the region of interest.
A resonant ($\delta=0$)  Gaussian solution to Eq.~(\ref{gaugeParax2}) is
displayed in Fig.~\ref{EBfieldFig}(a). The center of the Gaussian beam is 
shifted by an amount $x_\text{ctr} =F z^2/{2k} $, which is equivalent to
the motion of a charged particle in a constant electric field.

The control field phase profile (\ref{thetaEF}) can be implemented using, for
example, a phase plate. The assumption that the control fields do not depend
on $z$ implies that the Fresnel number for these fields must be above 1,
i.e. that the characteristic transverse distance over which these fields
significantly change must be larger than $\sim\sqrt{\lambda L}$, where $L$ is
the EIT cell length. This imposes a limitation on the magnitude of the
electric quasi-field: from \eeqref{thetaEF} we find $F\lesssim
\lambda^{-1/2}L^{-3/2}$ and thus $x_\text{ctr} \lesssim
\sqrt{L/\lambda}$. Assuming that the signal field also has a Fresnel number of
at least 1, and thus satisfies $2z_R\gtrsim L$ (with
Rayleigh length $z_R = k w^2/2$, $w$ being the signal beam width at the cell entrance), we find that in a realistic
experiment, the maximum possible signal beam displacement due to the
quasi-electric field is on the order of the signal beam width $w$.

To generate a homogeneous magnetic quasi-field along the z-axis the quantity
$B = \nabla\times A_{QQ}=2 \nabla_\perp \theta \times \nabla_\perp R$ should be constant.
However, it seems difficult to simultaneously achieve a vanishing
electric quasi-field ${\bf E} = -\nabla_\perp \Phi$. A choice that
minimizes the electric quasi-field around the origin is given by
$\theta = \sqrt{B/2}\, x$ and $R=\sqrt{B/2}\, y$. The quasi-potentials
then become $A_{QQ} = -B\, y\, {\bf e}_x$, which corresponds to the
Landau gauge in standard electrodynamics, and
$\Phi = B + 2 B^3 y^4 + O(y^6)$. If $\Phi$ is neglected, a Gaussian solution
to the paraxial wave equation is given by
\begin{eqnarray}\nonumber
   b_Q &=&\mathcal{N}
   \csc u(z)
   \exp \big[i\big(  {\textstyle \frac{B}{4}  }\cot u(z) \Delta{\bf x}^2+ \Delta {\bf x}\cdot {\bf p}_c    \\
   && -{\textstyle \frac{B}{2}  }\Delta x \Delta y -{\textstyle\frac{1}{2B}}{p_{c,x} p_{c,y}}
   + {\textstyle \frac{1}{2}  }y_c p_{c,y}\big)\big] \label{eq:magnSol}
\end{eqnarray}
where we have set $\Delta{\bf x} \equiv (x-x_c,y-y_c)$,
$u(z) \equiv Bz/(2k) -i \tanh^{-1} (2\eta)$ and $\eta \equiv B w^2/4$.
Here ${\bf x}_c = {\bf x}_0 +
(k/B)( {\bf x}'_0 \sin(Bz/k) + \tilde{{\bf x}}'_0
(1-\cos(Bz/k) )$ denotes the classical spiral
trajectory of a charged particle in a magnetic field, with
${\bf x}_c' = d {\bf x}_c/dz$,
initial position ${\bf x}_0$ and initial velocity
${\bf x}'_0$. For convenience we also have defined
$\tilde{{\bf x}}'_0 = (y'_0,-x'_0)$ and the classical canonical
momentum ${\bf p}_c$. We remark that $p_{c,x}$ is a constant of motion. 
The evolution of the signal mode is displayed in Fig.~2(b).

A surprising feature of solution (\ref{eq:magnSol}) is that the
diffractive divergence of the signal beam is reduced: the width squared of the Gaussian,
\begin{equation}
  \frac{2}{\text{Re}(i B \cot u)} = \frac{w^2}{8\eta^2}
  \left (
    1+ 4\eta^2 -(1-4\eta^2) \cos(\frac{Bz}{k})
  \right )\, ,
\end{equation}
varies periodically with $z$ instead of monotonically increasing.
This effect is known for electron wavepackets
\cite{takagi:JJAP2001} and can be understood as a consequence of the
circular motion of particles in a magnetic field: instead of dispersing,
two-dimensional particles in a magnetic field will simply move
on circles of different size (depending on their velocity), but
with the same angular velocity. The particle
cloud will therefore not spread but ``breathe''.

It remains to show that non-adiabatic coupling to other modes can be
suppressed for realistic experimental parameters. This is the case if
the strength of the gauge field terms coupling $b_Q$ to
other modes, which for the quasi-magnetic field are of the order $B/(2k)$,
are much smaller than the difference in the respective
linear susceptibilities $\chi_1$.
For the EIT mode $b_Q$, $\chi_1 = \delta/v_\text{EIT} $ with $v_\text{EIT}$ defined
above Eq.~(\ref{gaugeParax2}); for the other modes it can be
approximated by the susceptibility of a two-level medium,
$\chi_1 = -4 Ng^2(\delta-i\gamma/2)/(c \gamma^2)$. Evaluating this relation
at resonance leads to the condition $\eta \ll (k w)^2 n 3\pi/2$,
with $n\equiv N/(V k^3)$ being the number of atoms in the volume $k^{-3}$,
which can easily be fulfilled in an experiment.

{\em Aharonov-Bohm potential for photons.---}
One of the most intriguing phenomena of charged quantum particles
in electromagnetic fields is the Aharonov-Bohm (AB) effect
\cite{aharonov59}. Its two astonishing features are (i)
a phase shift induced by the vector potential in a region
in which electric and magnetic fields are absent, and (ii)
its topological nature: the phase shift does
not depend on the particle trajectory as long as it encloses
a magnetic flux.
Because (unlike genuine electromagnetism) the potential (\ref{AW}) is
a differential function of the control fields, it is impossible to
simulate feature (i) with quasi-charged
photons. However, we will show here that a mathematically
equivalent topological phase shift does exist for the optical case.

To generate an AB potential for photons we propose to use
two counter-rotating Laguerre-Gaussian control fields, i.e.,
fields that possess an orbital angular momentum.
If these control fields are spatially wider than the signal fields,
the corresponding Rabi frequencies can be approximated in cylindrical coordinates $(r,\varphi)$ by
$\Omega_1 = g_1 s_1 r e^{i\varphi}$ and $\Omega_2 = g_2 s_2 r e^{-i\varphi}$.
The gauge potentials (\ref{AQQ2}) then become
$A_{QQ} = -2R/r \; \vec{e}_\varphi$ and $\Phi = (1-4R^2)/r^2$,
with $R=\frac{1}{2}(|s_1|^2 - |s_2|^2)/(|s_1|^2 + |s_2|^2)$.
The potential $A_{QQ}$ corresponds exactly to an Aharonov-Bohm potential for
charged particles
as it is created by a solenoid.

Solutions of the paraxial wave equation
(\ref{gaugeParax2}) can be found in cylindrical coordinates
by expanding the field mode as $b_Q = r^{-1/2}
\sum_{m\in \text{Z \hspace*{-2mm}Z}} B_m (z,r) \exp (i m \varphi)$.
Because of $\Omega \sim r$, the EIT group velocity 
can be written as $v_\text{EIT} = \tilde{v}\,  r^2$ with
$\tilde{v} \equiv c \sqrt{|s_1|^2 +|s_2|^2}/N$.
Exact solutions are given by  Bessel functions,
\begin{equation}
  B_m = e^{-i \kappa^2 z/(2k)}
  \left ( \alpha_m \sqrt{\kappa r} J_{\nu}(\kappa r)
      + \beta_m \sqrt{\kappa r} Y_{\nu}(\kappa r)
       \right )\; ,
\end{equation}
with $\nu = \sqrt{1 +m^2 + 4 R m -2 k \tilde{v} \delta}$.
For monochromatic signal fields this corresponds to a rotation of
the transverse mode structure. For $R=\pm 1/2$
the potential transfers a unit amount of angular momentum
to the signal light, but generally the amount can vary continuously
between $-\hbar$ and $\hbar$. Signal photons in
the EIT mode therefore form a two-dimensional bosonic quantum system
in an Aharonov-Bohm potential.

{\em Conclusion.---}
We showed that EIT in a multi-$\Lambda$ system can be used to generate a variety of
geometric effects on propagating signal pulses that mimic the
behavior of a charged particle in an electromagnetic field.
We found specific arrangements of two spatially inhomogeneous pump fields in a
double-$\Lambda$ system
which generate quasi-gauge potentials which correspond to
 constant electric and magnetic fields. Furthermore topological effects like
 the Aharonov-Bohm phase shift
can be induced. The latter
is significantly different from the proposal of Ref.~\cite{PRA60:4301} in that
it is based on spatially inhomogeneous pump fields rather than the Doppler
effect in moving media.

This paper investigated EIT in systems with two ground levels. In such a
system, there is only one EIT mode, which results in an Abelian U(1) gauge
theory, making the physics analogous to electromagnetism. By extending to
multiple ground levels, it may be possible to obtain multiple EIT modes and
model non-Abelian gauge potentials. This will be explored in a future
publication.
\acknowledgments

We thank David Feder and Alexis Morris for fruitful discussions.
This work was supported by iCORE, NSERC, CIAR, Quantum{\em Works} and CFI.


\begin{thebibliography}{10}
\bibitem{GaugeFootnote} Note that this gauge transformation acts on the EIT mode
$\hat b_Q$ only and is therefore different from the gauge transformation
discussed above.
\end{thebibliography}
\end{document}